\documentclass[sigconf, nonacm]{acmart}
\AtBeginDocument{%
  }

\acmYear{2026}

\usepackage{multirow}
\usepackage{tikz}
\usepackage{dsfont}
\usepackage{float}     
\usepackage{placeins}  
\usetikzlibrary{arrows.meta, positioning, calc, fit, shapes.geometric, shapes.misc, backgrounds}

\begin{document}

\title[Balancing Trial and Reorder: A Hybrid Sequential Transformer–GBDT Ranker for On-Demand Delivery]{Balancing Trial and Reorder: A Hybrid Sequential Transformer–GBDT Ranker for On-Demand Delivery}

\author{Marcel Kurovski}
\email{marcel.kurovski@wolt.com}
\orcid{0009-0004-9220-8737}
\affiliation{%
  \institution{Wolt (DoorDash, Inc.)}
  \city{Munich}
  \country{Germany}
}

\author{Attila Nagy}
\email{attila.nagy@wolt.com}
\orcid{0000-0002-2110-8681}
\affiliation{%
  \institution{Wolt (DoorDash, Inc.)}
  \city{Munich}
  \country{Germany}
}

\author{Steffen Klempau}
\email{steffen.klempau@wolt.com}
\orcid{0009-0002-1187-0100}
\affiliation{%
  \institution{Wolt (DoorDash, Inc.)}
  \city{Berlin}
  \country{Germany}
}

\author{Aleksandr Fedintsev}
\email{aleksandr.fedintsev@wolt.com}
\orcid{0000-0002-9951-8322}
\affiliation{%
  \institution{Wolt (DoorDash, Inc.)}
  \city{Helsinki}
  \country{Finland}
}



\begin{abstract}
On a delivery platform, personalized store ranking greatly influences what
users find and order. Unlike digital-only domains, candidate stores are
local and bound by real-time availability and delivery operations. One
central modeling tension is between surfacing new stores for trial and
preserving ranking quality for sessions with reorder intent.

We present \emph{Universal Venue Ranker}~(UVR), a production system
deployed at Wolt that pairs a bidirectional transformer encoder for
sequential user modeling with a GBDT ranker integrating contextual, user,
and store features. Trained across all stores and domains of a country
while enforcing local delivery constraints at inference, UVR replaces four
previously separate ranking models---three for restaurants, one for
retail---with a single unified system.
Label smoothing and trial-biased sample weighting steer the model toward
new stores, lifting offline trial MRR by $+12\%$ to $+30\%$ over
production while regressing reorder MRR in five of six countries. These regressions leave Global CVR---our core online metric, which
blends trial and reorder sessions---statistically unchanged.

We validate UVR in three consecutive A/B tests, the first two across Wolt's largest operating markets and the
third spanning all operating countries and both domains. UVR V1 delivers
$+5.5\%$ Merchant Trial Rate and $+0.16\%$ Global CVR over the previous
production ranker; V2 adds a further $+0.45\%$ Merchant Trial Rate on top;
and V3, our cross-domain unification of the restaurant and retail rankers,
adds a further $+1.31\%$ Retail Merchant Trial Rate, together accounting
for substantial incremental gross order value and a materially simplified
serving stack.
\end{abstract}




\maketitle

\section{Introduction}
\label{sec:introduction}

Wolt, a DoorDash company, is a local commerce platform operating across 30+ countries and more than 1,000 cities in Europe and Asia for on-demand delivery across restaurants, grocery, and retail. Together with DoorDash, the group serves over 56 million monthly active users across 40+ countries.\footnote{As of December 2025}
Users access this selection through multiple surfaces, from multi-list (carousel) surfaces to single ranked lists~\cite{rahdari2022magic}, making personalized store ranking the central task that determines which stores users see, click, and purchase from. Throughout this paper we use \emph{store} as the umbrella term covering restaurants, grocery, and retail stores.

Recommender systems (RS) on delivery platforms face challenges absent in digital-only domains such as streaming platforms or social networks.
Candidate stores are local: eligibility depends on delivery location, distance constraints, real-time courier availability, merchant operating hours and workload, rendering the stores candidate set dynamic and sparse at every request.
As a multi-sided marketplace, rankings must balance consumer relevance, merchant exposure, and platform economics (objectives often in tension~\cite{mehrotra2019recommendations, zheng2019multi}) while generalizing across heterogeneous verticals (restaurants, grocery, and retail), each with distinct consumption patterns and business constraints.
A further tension is between \emph{reorder} and \emph{trial} purchases--the delivery-domain manifestation of the explore--exploit trade-off~\cite{chen2021values}: users return to trusted stores out of habit or convenience, yet also expect to find stores they have not tried before, and the ranker must actively balance both. We operationalize this trade-off as a constrained multi-objective problem, implemented via empirically-tuned per-session weighting (Section~\ref{sec:training-ranker}).

User behavior is sequential in nature, and purchase histories encode strong predictive signals for future decisions~\cite{sun2019bert4rec, pancha2022pinnerformer, tran2024pisa}.

We present \emph{Universal Venue Ranker}~(UVR), a production system that pairs a bidirectional transformer encoder for sequential user modeling with a gradient-boosted decision-tree (GBDT) ranker integrating contextual and content features.
Trained per country across all stores and domains, UVR replaces the previous production ranking stack with a single unified model.
We evaluate UVR on offline benchmarks and in large-scale A/B tests, observing improvements in conversion rate and trial rate, which translate into increased gross order value. We present this as a real-world industry case study of session-aware, cross-domain personalization deployed at production scale across 30+ countries.

\begin{figure}[t]
    \centering
    \includegraphics[width=0.5\columnwidth]{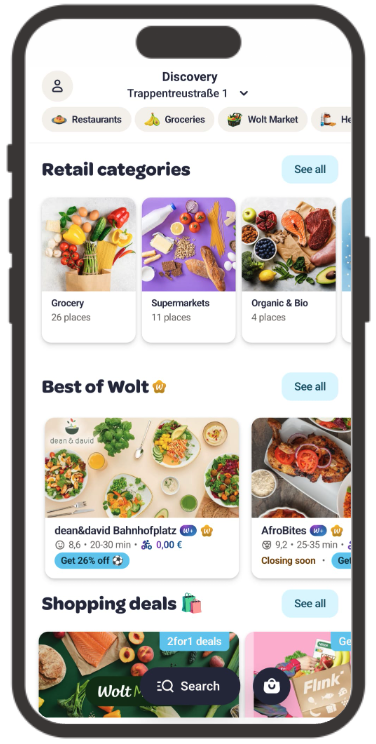}
    \caption{Wolt's Discovery Page}
    \label{fig:discovery-page}
\end{figure}

\section{Related Work}
\label{sec:related-work}
\paragraph{Transformer-based Models for Sequential User Modeling.}
Causal-attention (SASRec~\cite{kang2018sasrec}) and bidirectional-attention
(BERT4Rec~\cite{sun2019bert4rec}) transformer architectures superseded earlier Markov-chain and RNN-based sequential recommenders, with subsequent work advancing training objectives and sampling strategies~\cite{petrov2022replicability, petrov2023gsasrec}. They have been adopted across multiple industries; Tran et al. provide a comprehensive survey of transformers in recommendation \cite{tran2024pisa}.
Transformers have further proven effective in combination with \emph{Semantic IDs}
for retrieval and search~\cite{rajput2023recommender, penha2025semantic,
he2026plum}.

\paragraph{Recommender Systems in the Delivery Domain.}
Food and delivery recommendation has been studied through benchmark
datasets~\cite{assylbekov2023dhrd, wang2025meituan}, food-specific
methods~\cite{li2024food}, and production
architectures~\cite{egg2021online, meituan2021dual, feng2024context,
meituan2025longterm, han2025mtgr}. Relative to general e-commerce, delivery
platforms introduce compounding constraints such as local candidate sets, real-time
availability, and multi-stakeholder objectives, which remain underexplored in
the RS literature. MTGR~\cite{han2025mtgr} exemplifies a recent industrial trend
toward target-aware, generative multi-candidate architectures that score all
candidates within a single model pass. UVR's classifier shares this
\emph{multi-candidate, single-inference} property--one forward pass per request
scores the entire candidate slate (Section~\ref{sec:training-ranker})--but via a
simpler, non-target-aware bidirectional encoder rather than a generative
target-aware one.

\paragraph{Reorder vs. Trial Behavior.}
The tension between reorder and trial behavior has been studied through
surrogate metrics for long-term engagement~\cite{wang2022surrogate} and frameworks
for balancing trial and reorder behavior~\cite{spotify2020diversity, chen2021values,
su2024long, tran2024pisa}, primarily in music and video streaming. In on-demand
delivery, users similarly oscillate between reorder from trusted stores and trial of
new ones~\cite{li2024food}, and ranking must actively facilitate both.

\paragraph{Cross-domain Recommender Systems.}
Cross-domain recommender systems~(CDRS) transfer knowledge across domains to
mitigate cold-start and improve recommendation quality~\cite{nazari2020spotifycdrs, elkahky2015cross}. Our
multi-domain setting calls for this directly: a unified model lets behavior learned
in one vertical inform ranking in another, and naturally handles user cold-start
across domains.

UVR integrates these four threads into a single production system.

\section{Problem Formulation}
\label{sec:problem-formulation}

We frame personalized store ranking as a sequential recommendation
problem with explicit time conditioning. Let $\mathcal{U}$ denote
the set of users and $\mathcal{I}$ the set of stores. For each user
$u \in \mathcal{U}$, let
\[
  S_u \;=\; \bigl(\,i_1,\, i_2,\, \dots,\, i_{n_u}\,\bigr),
  \qquad i_k \in \mathcal{I},
\]
denote their lifetime purchase sequence ordered by ascending
timestamp, with $t_k^{u}$ the absolute timestamp of the $k$-th
purchase.

For any reference time $t$, we further introduce the time-truncated
purchase subsequence $S_u^{<t}$ (ordered) and the corresponding
store set $\mathcal{I}_u^{<t}$ (unordered):
\begin{align*}
  S_u^{<t}        &= \bigl(i_k \in S_u \,:\, t_k^{u} < t\bigr), \\
  \mathcal{I}_u^{<t} &= \{\, i \in \mathcal{I} \,:\, i \in S_u^{<t}\,\}.
\end{align*}

At request time, given a user $u$ at delivery location $l$ and
time $t$, the ranker receives a candidate set
$\mathcal{I}_{u,l}^{t} \subseteq \mathcal{I}$ determined by
vicinity and operational constraints.\footnote{The retrieval of
$\mathcal{I}_{u,l}^{t}$ from $\mathcal{I}$ is out of scope for
this work; in our setting it is driven by factors such as the
size of the delivery area, store availability, and courier
availability.} The task is to impose a ranking on
$\mathcal{I}_{u,l}^{t}$ via a scoring function
\[
  g\bigl(i \,\big|\, u, l, t\bigr) \in \mathbb{R}, \quad i \in \mathcal{I}_{u,l}^{t},
\]
such that descending order by $g$ minimizes the rank of the store
the user purchases from.

We denote by $r_{u,i}^{t} \in \{0,1\}$ the indicator that user $u$
purchases from store $i$ at time $t$\footnote{We use the purchase
signal as our primary target; clicks enter
Section~\ref{sec:ranker} as an auxiliary indicator
$c_{u,i,l}^{t}$. The framework extends naturally to further
interaction types (impressions, dwell time, \dots).}. We assume single-purchase sessions:
\begin{equation}
  \sum_{i \in \mathcal{I}_{u,l}^{t}} r_{u,i}^{t} \;=\; 1.
  \label{eq:single-purchase}
\end{equation}
Sessions with multiple purchases are reduced to the store with the
smallest production rank that received a purchase; sessions without
any purchase are discarded.

We further partition purchases into three types according to a
user's prior history:
\begin{equation}
  \mathrm{type}(u, i, t) \;=\;
  \begin{cases}
    \mathrm{\textsc{cs}}  & \text{if } r_{u,i}^{t} = 1 \text{ and } \mathcal{I}_u^{<t} = \emptyset, \\
    \mathrm{\textsc{new}} & \text{if } r_{u,i}^{t} = 1 \text{ and } i \notin \mathcal{I}_u^{<t}, \\
    \mathrm{\textsc{rec}} & \text{if } r_{u,i}^{t} = 1 \text{ and } i \in \mathcal{I}_u^{<t}.
  \end{cases}
  \label{eq:purchase-type}
\end{equation}
Cold-start (\textsc{cs}) purchases are from users with no
prior purchase history; \textsc{new} and \textsc{rec} distinguish a
trial order (first order from a store) from a recurring order (reorder) among returning users.

\begin{table}[t]
  \caption{Notation overview.}
  \label{tab:notation}
  \small
  \begin{tabular}{ll}
    \toprule
    Symbol & Meaning \\
    \midrule
    $\mathcal{U}$, $\mathcal{I}$ & user / store sets \\
    $u, i$ & a user / a store \\
    $l, t$ & delivery location, absolute timestamp \\
    $S_u$ & lifetime purchase sequence of user $u$ \\
    $S_u^{<t}$ & subsequence of $S_u$ with $t_k^{u} < t$ \\
    $\mathcal{I}_u^{<t}$ & set of distinct stores in $S_u^{<t}$ \\
    $\mathcal{I}_{u,l}^{t}$ & stores available to $u$ at $(l,t)$ \\
    $r_{u,i}^{t}$ & purchase indicator, $\in \{0,1\}$ \\
    $\mathrm{type}(u,i,t)$ & purchase type $\in \{\mathrm{\textsc{cs}},\mathrm{\textsc{new}},\mathrm{\textsc{rec}}\}$ \\
    $h_i(u, l, t)$ & transformer (classifier) score for store $i$ \\
    $g(i \mid u, l, t)$ & GBDT (ranker) score for store $i$ \\
    $\mathbf{x}_{u,i,l}^{t}$ & ranker feature vector \\
    \bottomrule
  \end{tabular}
\end{table}

\section{Methodology}
\label{sec:methodology}

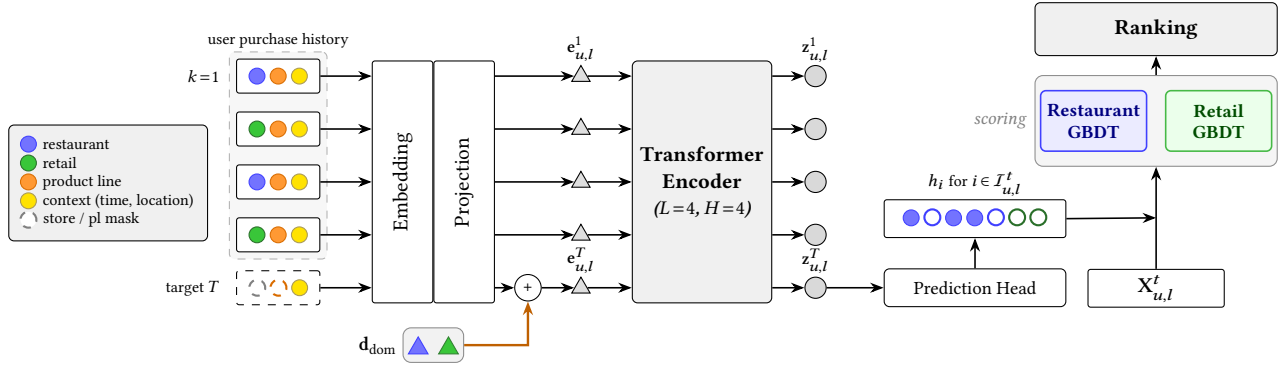
\begin{figure*}[t]
  \centering
  \begin{tikzpicture}[
    font=\small,
    >={Stealth[length=1.8mm]},
    crest/.style={fill=blue!55, draw=blue!75},
    crestmsk/.style={draw=blue!75, fill=white, thick},
    cretail/.style={fill=green!55!gray, draw=green!45!gray!60!black},
    cretailmsk/.style={draw=green!45!gray!60!black, fill=white, thick},
    cpl/.style={fill=orange!80, draw=orange!85!black},
    cplmsk/.style={draw=orange!85!black, fill=white, dashed, thick},
    cctx/.style={fill=yellow!85!orange, draw=yellow!75!orange!60!black, solid},
    dot/.style={circle, inner sep=0pt, minimum size=2.1mm},
    token/.style={draw, rounded corners=1pt, minimum width=11mm,
                  minimum height=4.5mm, inner sep=1pt, fill=white},
    tokenT/.style={draw, dashed, rounded corners=1pt, minimum width=11mm,
                   minimum height=4.5mm, inner sep=1pt, fill=white},
    layerbox/.style={draw, rounded corners=1pt, fill=white, align=center,
                     inner sep=2pt},
    encbox/.style={draw, rounded corners=2pt, fill=gray!12, align=center,
                   inner sep=3pt, font=\small\bfseries},
    headbox/.style={draw, rounded corners=2pt, fill=white, align=center,
                    inner sep=2pt, font=\footnotesize},
    gbdtrest/.style={draw=blue!75, line width=0.6pt, fill=blue!8,
                     rounded corners=2pt, align=center,
                     font=\bfseries\footnotesize, text=blue!50!black,
                     inner sep=2pt, minimum width=12mm, minimum height=8mm},
    gbdtretail/.style={draw=green!45!gray, line width=0.6pt, fill=green!8,
                       rounded corners=2pt, align=center,
                       font=\bfseries\footnotesize, text=green!30!black,
                       inner sep=2pt, minimum width=12mm, minimum height=8mm},
    rankbox/.style={draw, rounded corners=2pt, fill=gray!12, align=center,
                    font=\bfseries\small, inner sep=2pt,
                    minimum width=30mm, minimum height=7mm},
    scoringwrap/.style={draw=black!45, fill=gray!8, rounded corners=3pt,
                        line width=0.4pt, minimum width=30mm, minimum height=11mm},
    domwrap/.style={draw=black!45, fill=gray!10, rounded corners=3pt,
                    line width=0.4pt},
    plusnode/.style={circle, draw, fill=white, inner sep=0pt, minimum size=3.5mm,
                     font=\tiny},
    tri/.style={regular polygon, regular polygon sides=3, draw, fill=gray!25,
                inner sep=0pt, minimum size=2.8mm},
    trirest/.style={regular polygon, regular polygon sides=3, draw=blue!75,
                    fill=blue!55, inner sep=0pt, minimum size=3mm},
    triretail/.style={regular polygon, regular polygon sides=3,
                      draw=green!45!gray!70!black, fill=green!55!gray,
                      inner sep=0pt, minimum size=3mm},
    zdot/.style={circle, draw, fill=gray!30, inner sep=0pt, minimum size=2.8mm},
    arr/.style={->, semithick},
    dombus/.style={draw=orange!75!black, line width=0.9pt},
    leglabel/.style={font=\scriptsize, anchor=west, inner sep=1pt},
  ]

  \def\ya{4.0}\def\yb{3.3}\def\yc{2.6}\def\yd{1.9}\def\yT{1.2}

  \begin{scope}[xshift=1.5cm]

  \node[token] (tk1) at (0, \ya) {%
    \tikz[baseline=-1pt]{
      \node[dot, crest]   at (-0.28, 0) {};
      \node[dot, cpl]     at ( 0,    0) {};
      \node[dot, cctx]    at ( 0.28, 0) {};}};
  \node[token] (tk2) at (0, \yb) {%
    \tikz[baseline=-1pt]{
      \node[dot, cretail] at (-0.28, 0) {};
      \node[dot, cpl]     at ( 0,    0) {};
      \node[dot, cctx]    at ( 0.28, 0) {};}};
  \node[token] (tk3) at (0, \yc) {%
    \tikz[baseline=-1pt]{
      \node[dot, crest]   at (-0.28, 0) {};
      \node[dot, cpl]     at ( 0,    0) {};
      \node[dot, cctx]    at ( 0.28, 0) {};}};
  \node[token] (tk4) at (0, \yd) {%
    \tikz[baseline=-1pt]{
      \node[dot, cretail] at (-0.28, 0) {};
      \node[dot, cpl]     at ( 0,    0) {};
      \node[dot, cctx]    at ( 0.28, 0) {};}};
  \node[tokenT] (tkT) at (0, \yT) {%
    \tikz[baseline=-1pt]{
      \node[dot, draw=black!50, fill=white, dashed, thick] at (-0.28, 0) {};
      \node[dot, cplmsk]                                    at ( 0,    0) {};
      \node[dot, cctx]                                      at ( 0.28, 0) {};}};
  \node[font=\scriptsize, anchor=east, xshift=-3pt] at (tk1.west) {\,$k\!=\!1$};
  \node[font=\scriptsize, anchor=east, xshift=-3pt] at (tkT.west) {\,target $T$};

  \begin{scope}[on background layer]
    \node[draw=black!30, dashed, rounded corners=2pt, fill=black!3,
          fit=(tk1) (tk2) (tk3) (tk4), inner sep=2.5pt,
          label={[font=\scriptsize, align=center, inner sep=1pt]above:user purchase history}]
      {};
  \end{scope}

  \node[layerbox, minimum width=8mm, minimum height=32mm, anchor=east]
    (emb) at (2.05, 2.6) {\rotatebox{90}{Embedding}};
  \node[layerbox, minimum width=8mm, minimum height=32mm, anchor=west]
    (prj) at (2.05, 2.6) {\rotatebox{90}{Projection}};
  \foreach \k in {1,2,3,4,T} {
    \draw[arr] (tk\k.east) -- (tk\k.east -| emb.west);
  }

  \node[plusnode] (epT) at (3.3, \yT) {$+$};
  \draw[arr] (epT.west -| prj.east) -- (epT.west);

  \foreach \k/\y in {1/\ya, 2/\yb, 3/\yc, 4/\yd} {
    \node[tri] (e\k) at (4.0, \y) {};
    \draw[arr] (e\k.west -| prj.east) -- (e\k.west);
  }
  \node[tri] (eT) at (4.0, \yT) {};
  \draw[arr] (epT.east) -- (eT.west);
  \node[font=\scriptsize, anchor=south, inner sep=1pt] at (e1.north) {$\mathbf{e}_{u,l}^{1}$};
  \node[font=\scriptsize, anchor=south, inner sep=1pt] at (eT.north) {$\mathbf{e}_{u,l}^{T}$};

  \node[trirest, anchor=center]   (dr) at (1.85, 0.4) {};
  \node[triretail, anchor=center] (dt) at (2.25, 0.4) {};
  \begin{scope}[on background layer]
    \node[domwrap, fit=(dr) (dt), inner sep=3pt,
          label={[font=\footnotesize, inner sep=2pt]left:$\mathbf{d}_{\mathrm{dom}}$}]
      (dom) {};
    \draw[->, draw=orange!75!black, line width=0.9pt]
      (dom.east) -| (epT.south);
  \end{scope}

  \node[encbox, minimum width=18mm, minimum height=32mm, anchor=center]
    (enc) at (5.6, 2.6) {Transformer\\Encoder\\{\footnotesize\itshape($L\!=\!4$, $H\!=\!4$)}};
  \foreach \k/\y in {1/\ya, 2/\yb, 3/\yc, 4/\yd, T/\yT} {
    \draw[arr] (e\k.east) -- (e\k.east -| enc.west);
  }

  \foreach \k/\y in {1/\ya, 2/\yb, 3/\yc, 4/\yd, T/\yT} {
    \node[zdot] (z\k) at (7.1, \y) {};
    \draw[arr] (z\k.west -| enc.east) -- (z\k.west);
  }
  \node[font=\scriptsize, anchor=south, inner sep=1pt] at (z1.north) {$\mathbf{z}_{u,l}^{1}$};
  \node[font=\scriptsize, anchor=south, inner sep=1pt] at (zT.north) {$\mathbf{z}_{u,l}^{T}$};

  \node[headbox, anchor=west, minimum width=24mm, minimum height=5mm,
        font=\footnotesize] (head) at (8.0, \yT) {Prediction Head};
  \draw[arr] (zT.east) -- (head.west);


  \node[layerbox, anchor=south, minimum height=5mm, minimum width=24mm] (scores)
    at ([yshift=4mm]head.north) {%
      \tikz[baseline=-1pt]{
        \node[dot, crest]      at (0.22, 0) {};
        \node[dot, crestmsk]   at (0.50, 0) {};
        \node[dot, crest]      at (0.78, 0) {};
        \node[dot, crest]      at (1.06, 0) {};
        \node[dot, crestmsk]   at (1.34, 0) {};
        \node[dot, cretailmsk] at (1.62, 0) {};
        \node[dot, cretailmsk] at (1.90, 0) {};
      }};
  \node[font=\scriptsize, anchor=south, inner sep=1pt] at (scores.north)
    {$h_i$ for $i\!\in\!\mathcal{I}_{u,l}^{t}$};
  \draw[arr] (head.north) -- (scores.south);

  \node[scoringwrap, anchor=south, minimum width=32mm, minimum height=12mm]
    (scoring) at (11.6, 2.80) {};

  \node[layerbox, anchor=north, minimum width=18mm, minimum height=5mm]
    (feats) at ([yshift=-13.5mm]scoring.south) {$\mathbf{X}_{u,l}^{t}$};

  \node[gbdtrest, anchor=west, minimum width=14mm, minimum height=8mm]
    (gr) at ([xshift=2pt]scoring.west) {Restaurant\\GBDT};
  \node[gbdtretail, anchor=east, minimum width=14mm, minimum height=8mm]
    (gt) at ([xshift=-2pt]scoring.east) {Retail\\GBDT};
  \node[font=\itshape\footnotesize, gray, anchor=east, inner sep=3pt]
    at (scoring.west) {scoring};

  \coordinate (xsmerge) at ($(feats.north)!0.5!(scoring.south)$);
  \draw[arr] (feats.north) -- (scoring.south);
  \draw[arr] (scores.east) -- (xsmerge);

  \node[rankbox, anchor=south, minimum width=32mm] (rank)
    at ([yshift=2.5mm]scoring.north) {Ranking};
  \draw[arr] (scoring.north) -- (rank.south);

  \end{scope}  

  \def\lxL{-1.8}
  \node[dot, crest]                                    (lA1) at (\lxL, 3.10) {};
  \node[leglabel, anchor=west]                         (lL1) at (lA1.east)  {\,restaurant};
  \node[dot, cretail]                                  (lA2) at (\lxL, 2.85) {};
  \node[leglabel, anchor=west]                         (lL2) at (lA2.east)  {\,retail};
  \node[dot, cpl]                                      (lA3) at (\lxL, 2.60) {};
  \node[leglabel, anchor=west]                         (lL3) at (lA3.east)  {\,product line};
  \node[dot, cctx]                                     (lA4) at (\lxL, 2.35) {};
  \node[leglabel, anchor=west]                         (lL4) at (lA4.east)  {\,context (time, location)};
  \node[dot, draw=black!50, fill=white, dashed, thick] (lA5) at (\lxL, 2.10) {};
  \node[leglabel, anchor=west]                         (lL5) at (lA5.east)  {\,store / pl mask};
  \begin{scope}[on background layer]
    \node[draw, rounded corners=2pt, fill=gray!12, line width=0.4pt,
          fit=(lA1) (lL1) (lA5) (lL5) (lL4), inner sep=4pt] {};
  \end{scope}

  \end{tikzpicture}
  \caption{UVR architecture: bidirectional Transformer classifier $h$ feeding
  domain-specific GBDT rankers $g$.}
  \label{fig:uvr-architecture}
\end{figure*}

As depicted in Figure~\ref{fig:uvr-architecture}, UVR is a
two-stage system:
\begin{enumerate}
  \item a bidirectional transformer \emph{classifier} $h$ that
    maps a user's purchase sequence and request context to a
    vector of per-store scores (Section~\ref{sec:classifier});
  \item a GBDT \emph{ranker} $g$ that
    consumes $h$ as a single scalar feature alongside content,
    contextual, and user--store interaction features
    (Section~\ref{sec:ranker}).
\end{enumerate}
Concretely, the final score is
\begin{equation}
  g\bigl(i \,\big|\, u, l, t\bigr)
    \;=\; \mathrm{GBDT}\!\left(\, h_i(u, l, t),\;\mathbf{x}_{u,i,l}^{t} \,\right),
  \label{eq:uvr-coupling}
\end{equation}
where $h_i(u, l, t) \in \mathbb{R}$ is the raw transformer logit for
store $i$ and user $u$ at time $t$, and $\mathbf{x}_{u,i,l}^{t}$
collects content and contextual features. We pass the raw logit
rather than a softmax probability because the GBDT ranker is
invariant to monotonic feature transformations.

UVR's contribution is twofold: (i) it couples a sequential
transformer to a tree-based ranker inside a single deployed
stack, a stacking pattern established in production
search~\cite{yin2016ranking} and instantiated here under hard
local-availability constraints; and (ii) purchase-type-aware weighting makes the trial/reorder trade-off directly tunable (Section~\ref{sec:training-ranker}). The two stages are fit in sequence on disjoint
data slices. Both are realized
over a unified store vocabulary $\mathcal{I} =
\mathcal{I}^{\text{rest}} \cup \mathcal{I}^{\text{retail}}$,
covering restaurants and retail. The following sections describe
this cross-domain UVR architecture; the experimental study
(Section~\ref{sec:experiments}) reports on the restaurant-only
configurations (V1, V2) followed by the cross-domain variant
(V3), whose online A/B result is reported in
Section~\ref{sec:online}.

\subsection{Classifier: User Sequence Transformer for Next-Purchase Prediction}
\label{sec:classifier}

The classifier $h$ is an encoder-only
transformer~\cite{vaswani2017attention} trained as a multi-class
store classifier over $\mathcal{I}$, adopting the
bidirectional-attention masked-prediction backbone of
BERT4Rec~\cite{sun2019bert4rec} but narrowing its random-position
cloze objective to mask only the most recent purchase in $S_u$.

\textbf{Partial Target Masking.}
Three pieces of information are available to the model at inference time and are therefore supplied unmasked: (i) the request
timestamp $t$, from which we derive hour-of-day, day-of-week and
calendar week; (ii) the H3\footnote{H3 is an open-source hierarchical
hexagonal geospatial indexing system maintained by Uber that
partitions the surface of the Earth into nested hexagonal cells.
We use resolution~8, whose cells have an average area of
approximately $0.74\,\text{km}^{2}$, comparable to a small
neighborhood or a few city blocks. See \url{https://h3geo.org/}.} cell ID of the user's location $l$ (see
Section~\ref{sec:embedding-layer}); and (iii) the requested domain
(restaurant or retail), which is known because most user surfaces
are partitioned by domain. Two pieces of information are unknown
and are masked: (a) the target store ID $i^{\star}$, which is the
prediction target, and (b) the product line of $i^{\star}$, which
is masked because the model is queried per domain rather than per
product line; supplying the exact product line at the target
position would constitute label leakage.

\subsubsection{Embedding Layer}
\label{sec:embedding-layer}

Each step of the input sequence is described by six categorical
features and one continuous feature, summarized in
Table~\ref{tab:per-step-features}. The categoricals are: target
store ID $\in \mathcal{I} \cup \{\textsc{mask}\}$; product line
$\in \text{PL} \cup \{\textsc{mask}^{\text{rest}}, \textsc{mask}^{\text{retail}}\}$,
where the two domain-specific mask tokens are appended beyond the
real product-line vocabulary; week-of-year; day-of-week;
hour-of-day; and the H3 cell ID of the user location. We deliberately do not feed raw latitude and
longitude into the model; discretizing location to an H3 cell ID aligns with the granularity of our delivery operations and shares parameters across all requests originating from the same cell, which may also benefit cold-start users.

The continuous feature is the time gap
$\Delta_k^{u} = t - t_k^{u}$ from each historical position to the
target. We transform $\Delta_k^{u}$ by a log-affine map for
numerical stability:
\begin{equation}
  \tilde{\Delta}_k^{u} \;=\; \log\bigl(\alpha\,\Delta_k^{u} + \beta\bigr),
  \qquad \alpha, \beta > 0.
  \label{eq:diff-transform}
\end{equation}
At the target step (where $k = T$ and $\Delta_T^{u} = 0$),
$\tilde{\Delta}_T^{u} = \log \beta$ acts as a constant offset.

\begin{table}[t]
  \caption{Features Representing Purchases in User Sequences}
  \label{tab:per-step-features}
  \small
  \begin{tabular}{lll}
    \toprule
    Feature & Source & Masked at target? \\
    \midrule
    store ID & past or target purchase & yes \\
    product line & past or target purchase & yes (per domain) \\
    H3 hexagon ID & delivery location $l$ & no \\
    hour-of-day & timestamp $t$ & no \\
    day-of-week & timestamp $t$ & no \\
    week-of-year & timestamp $t$ & no \\
    $\tilde\Delta_k^{u}$ & $t - t_k^{u}$ & no ($=\log\beta$ at target) \\
    \bottomrule
  \end{tabular}
\end{table}

Each categorical feature has its own learned embedding
function. Writing
$\mathcal{C} = \{\text{store}, \text{pl}, \text{hex}, \text{hour}, \text{dow}, \text{week}\}$
for the set of categorical fields, we associate one
embedding function per field,
\[
  E_c \colon \mathcal{X}_c \to \mathbb{R}^{d_c},
  \qquad c \in \mathcal{C},
\]
where $\mathcal{X}_c$ is the input vocabulary for field $c$
(for example,
$\mathcal{X}_{\text{store}} = \mathcal{I} \cup \{\textsc{mask}\}$
and
$\mathcal{X}_{\text{pl}} = \text{PL} \cup \{\textsc{mask}^{\text{rest}}, \textsc{mask}^{\text{retail}}\}$).
We denote by $x_k^c$ the value of field $c$ at step $k$.
The input representation at step $k$ is obtained by
concatenating the six per-field embeddings with the transformed
continuous gap and projecting to the model dimension
$d_{\text{model}}$:
\begin{equation}
  \mathbf{e}_k \;=\; W_{\text{in}} \Bigl[\,
    \bigl(E_c(x_k^c)\bigr)_{c \in \mathcal{C}}
    \,\Vert\, \tilde{\Delta}_k^{u}
  \,\Bigr] \;+\; \mathbf{b}_{\text{in}}
  \;+\; \mathds{1}[k = T]\,\mathbf{d}_{\mathrm{dom}(i^{\star})},
  \label{eq:step-embedding}
\end{equation}
where $\Vert$ denotes concatenation,
$\bigl(E_c(x_k^c)\bigr)_{c \in \mathcal{C}}$ is the
concatenation of the per-field embeddings in a fixed canonical
order, $W_{\text{in}}, \mathbf{b}_{\text{in}}$ are the
parameters of the linear input projection, and the indicator
$\mathds{1}[k = T]$ restricts the domain embedding
$\mathbf{d}_{\mathrm{dom}(i^{\star})} \in \mathbb{R}^{d_{\text{model}}}$
to the target step.

\textbf{Two-level domain conditioning.}
We inject the requested domain at the target step at two points
of the embedding pipeline. \emph{Pre-projection}, the target's
product line carries either $\textsc{mask}^{\text{rest}}$ or
$\textsc{mask}^{\text{retail}}$, which is processed by the
product-line embedding alongside the other five categoricals.
\emph{Post-projection}, the additive term
$\mathds{1}[k=T]\,\mathbf{d}_{\mathrm{dom}(i^{\star})}$ in
Eq.~\ref{eq:step-embedding} is a dedicated domain embedding,
restricted by the indicator to the target step. At $k = T$ we
additionally substitute the store and product-line inputs by their
mask tokens, $x_T^{\text{store}} = \textsc{mask}$ and
$x_T^{\text{pl}} = \textsc{mask}^{\mathrm{dom}(i^{\star})}$.

We found that supplying the domain signal only through the
pre-projection mask led to severe dilution as the representation
propagated through successive encoder layers; the additive
post-projection embedding instead acts as a clean, isolated signal
in $d_{\text{model}}$-space and sharpens conditioning without
disrupting the rest of the encoder input. We also experimented
with FiLM-style feature-wise affine
modulation~\cite{perez2018film} of the encoder activations but
did not observe a benefit over the simpler additive scheme.

\subsubsection{Encoder Layer}\footnote{Reported values are for V1 and V2. V3 widens the architecture to $H = 6$,
$d_{\text{model}} = 192$, $d_{\text{ff}} = 384$ and trains
for $E = 12$ epochs (instead of the $E = 15$ used by V1 and
V2); the encoder depth $L = 4$ and all other hyperparameters
in Section~\ref{sec:training-validation} are unchanged.}

We stack $L = 4$ identical transformer encoder layers with
pre-LayerNorm~\cite{xiong2020prelayernorm}, multi-head
self-attention with $H = 4$ heads, and a feed-forward sub-layer of hidden size
$d_{\text{ff}} = 160$. The model dimension is $d_{\text{model}} = 128$.
Sequences shorter than the context length $T$ are left-padded with zero embeddings;
padding positions are additionally excluded from attention via a key-padding mask.
No causal mask is applied: the target step attends to all historical positions.
We denote the encoder output at the target step by
$\mathbf{z}_{u,l}^{T} \in \mathbb{R}^{d_{\text{model}}}$.

\subsubsection{Prediction Layer}

A single linear head with weight matrix
$W_{\text{out}} \in \mathbb{R}^{|\mathcal{I}| \times d_{\text{model}}}$
and bias $\mathbf{b}_{\text{out}} \in \mathbb{R}^{|\mathcal{I}|}$
projects $\mathbf{z}_{u,l}^{T}$ to a logit vector over the full store
vocabulary,
\[
  h(u, l, t) \;=\; W_{\text{out}}\,\mathbf{z}_{u,l}^{T} + \mathbf{b}_{\text{out}} \;\in\; \mathbb{R}^{|\mathcal{I}|},
\]
and we write $h_i(u, l, t) = \bigl[W_{\text{out}}\bigr]_{i,:}\,\mathbf{z}_{u,l}^{T} + b_{\text{out},i}$
for the per-store logit.

At training time we minimize a per-sample-weighted, label-smoothed
softmax cross-entropy loss whose denominator is restricted to the
in-domain vocabulary
$\mathcal{I}^{\mathrm{dom}(i^{\star})}$ of the ground-truth store,
preventing cross-domain leakage during gradient updates. With
label-smoothing parameter $\varepsilon \in [0,1)$, the smoothed
target distribution is
\[
  \tilde{y}_j \;=\; (1 - \varepsilon)\,\mathbf{1}[\,j = i^{\star}\,] \;+\; \frac{\varepsilon}{|\mathcal{I}^{\mathrm{dom}(i^{\star})}|},
  \qquad j \in \mathcal{I}^{\mathrm{dom}(i^{\star})},
\]
and the loss is
\begin{equation}
  \mathcal{L}(\theta) \;=\; -\!\!\sum_{(u, l, t, i^{\star}) \in \mathcal{D}}\!\!\!
    w_{u, i^{\star}}^{t} \!\!\!\!\sum_{j \in \mathcal{I}^{\mathrm{dom}(i^{\star})}}\!\!\!\!
      \tilde{y}_j\,
    \log \frac{\exp h_{j}(u, l, t)}{\sum_{k \in \mathcal{I}^{\mathrm{dom}(i^{\star})}} \exp h_k(u, l, t)}.
  \label{eq:cls-loss}
\end{equation}

The per-sample weights $w_{u, i^{\star}}^{t}$ are tied to the purchase
type defined in Equation~\ref{eq:purchase-type} and follow
the ordering
$w^{\textsc{new}} > w^{\textsc{cs}} > w^{\textsc{rec}}$,
biasing the gradient towards stores the user has not yet purchased
from while still rewarding correct predictions of returning
purchases. At inference time the classifier is queried once per
request with its $(u, l, t)$ context; only the raw logits
$h_i(u, l, t)$ for the candidate stores
$i \in \mathcal{I}_{u,l}^{t}$ are retained and passed to the
ranker.

\textbf{Cold-start handling.}
For cold-start users the
encoder input collapses to a single (partially masked) target step. The cold-start signal is therefore implicit in the input
length: with no historical positions to attend to, the model
produces a domain-, location- and time-conditioned prior over
stores from the target step alone. Note that the contextual
features at the target (hex, hour, day-of-week, week) are not
themselves sufficient to identify a user as cold-start, since
two co-located requests at the same timestamp from different
users (one cold-start, one returning) share those features
exactly; the discriminating signal is the absence of historical
positions, not the contents of the target step.

\subsection{Ranker: GBDT for Listwise Store Ranking}
\label{sec:ranker}

The ranker $g$ is a GBDT model implemented with
CatBoost~\cite{prokhorenkova2018catboost}, taking
per-(user, candidate-store) feature vectors
$\mathbf{x}_{u,i,l}^{t}$ over the candidate set
$\mathcal{I}_{u,l}^{t}$; training and label construction are
detailed in Section~\ref{sec:training-ranker}.

The feature vector $\mathbf{x}_{u,i,l}^{t}$ comprises four
groups:
\begin{itemize}
  \item \textbf{Classifier signal} (single scalar): the
    per-candidate transformer logit $h_i(u, l, t)$.
  \item \textbf{Store content features}: rating, conversion
    rate, retention rate, and assortment size (e.g.\ menu size
    for restaurants, in-stock item count for retail), etc.
  \item \textbf{Contextual features}: delivery distance,
    delivery price, and location-derived signals, etc.
  \item \textbf{User--store interaction features}: recent
    click counts, mean basket size, and trial ratio, etc.
\end{itemize}
Two domain-specific rankers are trained because the content and
contextual feature sets differ structurally between restaurant
and retail candidates (e.g.\ menu size is restaurant-only;
in-stock item count is retail-only).

\begin{table}[t]
  \caption{Comparison of UVR versions. V1, V2, and V3 are
  evaluated in Section~\ref{sec:experiments}.}
  \label{tab:uvr-versions}
  \small
  \setlength{\tabcolsep}{4pt}
  \begin{tabular}{@{}lp{1.4cm}p{2.4cm}p{3.0cm}@{}}
    \toprule
    Version & Domains & Ranker Loss & Notable additions\\
    \midrule
    V1 & restaurants & \textsc{PairLogit} (country-specific \texttt{max\_pairs}) & --- \\
    V2 & restaurants & \textsc{YetiRank-Pairwise:NDCG} & label smoothing; trial weighting \\
    V3 & restaurants + retail & \textsc{YetiRank-Pairwise:NDCG} & cross-domain coupling; product-line embedding \\
    \bottomrule
  \end{tabular}
\end{table}

\subsection{Coupling and Cross-Domain Transfer}

Cross-domain transfer enters through (a) the chronologically
interleaved sequence $S_u^{<t}$, (b) the per-position
product-line embedding, and (c) the two-level domain
conditioning at the target step (Section~\ref{sec:embedding-layer}).
A user's restaurant history thus informs their retail
next-store prediction and vice versa, and the per-domain
rankers inherit this transfer through the shared classifier
logit $h_i(u,l,t)$.

\subsection{Model Training and Validation}
\label{sec:training-validation}

We train one country-specific UVR model for each of the 30
countries Wolt operates in, and re-train it daily. This is a
deliberate trade-off between training complexity, resource
utilization, model degradation, and store cold-start. We do
not observe a clear performance trend over the first
$\sim$28 days post-training; the daily cadence is instead
driven by store cold-start: the store, H3 cell, and
product-line embedding tables
(Section~\ref{sec:embedding-layer}) are populated only for
values observed during training, making the classifier
\emph{transductive} with respect to these three vocabularies
but \emph{inductive} with respect to time (hour-of-day,
day-of-week, week-of-year embeddings are allocated
exhaustively at construction). Without daily retraining the
number of unseen stores accumulates rapidly and degrades
predictions. A complementary mitigation through supplementary
content-based store features is left for future work, e.g.\
by adopting DenseRec~\cite{lichtenberg2025denserec}.

We train and validate UVR in two stages: first the classifier
(Section~\ref{sec:training-classifier}), then the ranker
(Section~\ref{sec:training-ranker}), since the trained
classifier is needed to score the candidate slates that
constitute the ranker's training, validation, and test data.
The two stages operate on disjoint data slices: the
classifier is trained on prefixes
$S_u^{<(t - 2\mathrm{d})}$, i.e.\ the last purchase of any sequence is guaranteed to happen before the ranker's validation and test sessions. This prevents both
the ranker's validation day (which drives early stopping) and
its held-out test day from leaking into the classifier's
training set.

\subsubsection{Classifier}
\label{sec:training-classifier}

\paragraph{Data split.}
We adopt a user-based split of user purchase sequences into a
$90\%$ training set $\mathcal{D}_{\text{c}}^{\text{train}}$ and
a $10\%$ validation set $\mathcal{D}_{\text{c}}^{\text{val}}$.
Any sequence in $\mathcal{D}_{\text{c}}^{\text{val}}$ that
references a store ID, an H3 cell ID, or a product line
absent from $\mathcal{D}_{\text{c}}^{\text{train}}$ is moved
back to the training set; given the transductive nature of
the classifier with respect to these three feature fields
(Section~\ref{sec:training-validation}), this guarantees that
every entity observed at validation time also occurs in the
training corpus and therefore has an associated learned
embedding. Sequences are capped to the most recent $M$
purchases per user, since we observe diminishing returns from
longer histories, consistent with the findings of Pancha et
al.~\cite{pancha2022pinnerformer}. See appendix Section~\ref{app:user-sequence-analysis} for a detailed analysis.

\paragraph{Validation decomposition.}
We further partition $\mathcal{D}_{\text{c}}^{\text{val}}$ by the
type of the target purchase, defined in
Equation~\ref{eq:purchase-type}. Writing $i_T^{u}$ for the most
recent (target) purchase of user $u$ at time $t_T^{u}$,
\begin{align}
  \mathcal{D}_{\text{c}}^{\text{val}_{\tau}} \;&=\; \bigl\{\,S_u \in \mathcal{D}_{\text{c}}^{\text{val}} \,:\, \mathrm{type}(u, i_T^{u}, t_T^{u}) = \tau\,\bigr\}, \notag \\
  &\qquad \tau \in \{\mathrm{\textsc{cs}}, \mathrm{\textsc{new}}, \mathrm{\textsc{rec}}\}.
  \label{eq:val-decomp}
\end{align}
This decomposition lets us monitor cold-start, trial, and
reorder performance independently during training.

\paragraph{Optimization.}
We train the transformer for $E = 15$ epochs with a training
batch size of $256$ and a validation batch size of $1024$. We use
AdamW\cite{loshchilov2017decoupled} with learning rate $\eta = 10^{-3}$ and weight decay
$\omega = 0.15$, combined with a multi-step learning-rate
schedule that decays $\eta$ by a factor of $\gamma = 0.1$ at
epochs $\{4, 6, 7\}$. Training runs in mixed-precision
\texttt{bfloat16} autocast on a single A10G GPU, with the loss in
Equation~\ref{eq:cls-loss} computed at label-smoothing
$\varepsilon = 0.2$ and per-sample weights ordered
$w^{\textsc{new}} > w^{\textsc{cs}} > w^{\textsc{rec}}$.

\paragraph{Checkpoint selection.}
After each epoch we evaluate the model on the three validation
subsets in Equation~\ref{eq:val-decomp}. For each store $i$,
let $\mathcal{N}_i \subset \mathcal{I}$ be the set of
same-country stores within Haversine distance $D$ of $i$, $i$
itself included. For each $S_u$ we score the classifier on
$\mathcal{N}_{i_T^u}$, compute the rank of $i_T^u$, and
aggregate to mean reciprocal rank (MRR) across
$\mathcal{D}_{\text{c}}^{\text{val}_{\tau}}$. On the
\textsc{new} slice we further exclude $\mathcal{I}_u^{<t}$
from the scored set, so the target competes only against
stores the user has not yet purchased from. We retain the
checkpoint with the highest MRR on
$\mathcal{D}_{\text{c}}^{\text{val}_{\textsc{new}}}$, optimizing
explicitly for trial recommendations; the \textsc{rec} and
\textsc{cs} MRRs serve as guardrails.

\subsubsection{Ranker}
\label{sec:training-ranker}

\paragraph{Data.}
The ranker operates over user sessions rather than purchase
sequences. Following the assumptions in
Section~\ref{sec:problem-formulation}, we focus on
single-purchase sessions
(Equation~\ref{eq:single-purchase}); sessions without any
purchase are dropped as low-intent noise. A session in which
several stores received a purchase is normalized to a
single-purchase session by keeping
$r_{u,i}^{t} = 1$ for the purchased store with the smallest
production rank and setting $r_{u,j}^{t} \leftarrow 0$ for every
other purchased store $j \neq i$ in the same session, so the
single-purchase constraint of Equation~\ref{eq:single-purchase}
holds for every session in the resulting dataset. We define
the ranker dataset as a set of session tuples
\begin{equation}
  \mathcal{D}_{r} \;=\; \bigl\{\,\bigl(u, l, t,\, i^{+},\, \mathcal{I}_{u,l}^{t},\, \mathbf{X}_{u,l}^{t},\, \rho^{\mathrm{prod}},\, \mathbf{o}_{u,l}^{t},\, \mathbf{c}_{u,l}^{t}\bigr)\,\bigr\},
  \label{eq:ranker-dataset}
\end{equation}
where the session is uniquely identified by the
$(u, l, t)$ triple, $i^{+} \in \mathcal{I}_{u,l}^{t}$ is the
purchased store (the unique $i$ for which $r_{u,i}^{t} = 1$
after the normalization above),
$\mathbf{X}_{u,l}^{t} = (\mathbf{x}_{u,i,l}^{t})_{i \in \mathcal{I}_{u,l}^{t}}$
collects the per-candidate feature vectors,
$\rho^{\mathrm{prod}} : \mathcal{I}_{u,l}^{t} \to \{1, \dots, |\mathcal{I}_{u,l}^{t}|\}$
is the production ranking served at request time, and
$\mathbf{o}_{u,l}^{t}, \mathbf{c}_{u,l}^{t} \in \{0, 1\}^{|\mathcal{I}_{u,l}^{t}|}$
are per-candidate impression and click indicators
(\,$o_{u,i,l}^{t} = 1$ iff $i$ was shown to the user during the
session, $c_{u,i,l}^{t} = 1$ iff $i$ was clicked\,). At this
stage $\mathbf{x}_{u,i,l}^{t}$ contains only content,
contextual, and user--store interaction features; the
classifier signal $h_i(u, l, t)$ is appended later
(``Coupling with the classifier'' below).

\paragraph{Time-based split.}
We use the most recent $30$ days of sessions and adopt a
time-based train/validation/test split, disjoint from the
classifier's user-based split: the first $28$ days form
$\mathcal{D}_{r}^{\text{train}}$, the second-to-last day forms
$\mathcal{D}_{r}^{\text{val}}$ (used for early stopping), and
the last day is held out as $\mathcal{D}_{r}^{\text{test}}$.
The two-day cutoff applied to the classifier's training
horizon (Section~\ref{sec:training-validation}) is what
enforces the disjointness: where a training prefix $S_u^{<t'}$
allowed to terminate on the ranker's validation or test day,
the corresponding target purchase would be the same event
that the ranker uses as its label on
$\mathcal{D}_{r}^{\text{val}}$ or
$\mathcal{D}_{r}^{\text{test}}$, and the classifier logit
$h_{i^{+}}(u, l, t)$ subsequently fed to the ranker as a feature
would carry direct knowledge of the held-out target. By
training the classifier only on prefixes
$S_u^{<(t - 2\mathrm{d})}$ we ensure no such overlap can
occur and that
$h_i(u, l, t)$, evaluated on
$\mathcal{D}_{r}^{\text{val}}$ and
$\mathcal{D}_{r}^{\text{test}}$, is a true out-of-sample
prediction.

\paragraph{Training-set restriction.}
On $\mathcal{D}_{r}^{\text{train}}$ we restrict each session
to its impressed stores
$\{i \in \mathcal{I}_{u,l}^{t} : o_{u,i,l}^{t} = 1\}$. These
are harder negatives than stores the user never saw: the
production system already considered them attractive enough to
display, so the ranker must learn to discriminate purchases
from \emph{competitive} alternatives rather than from random
stores in the catalogue. On $\mathcal{D}_{r}^{\text{val}}$ and
$\mathcal{D}_{r}^{\text{test}}$ we keep the entire candidate
set $\mathcal{I}_{u,l}^{t}$, including unseen stores, because
this matches inference: at request time the ranker must order
the full slate without knowing which subset will be impressed
to the user. The training data is therefore drawn from
historical user session logs joined with impression events, while
validation and test data couple the same sessions with the
full backend slate.

\paragraph{Coupling with the classifier.}
For each session $(u, l, t, \dots) \in \mathcal{D}_{r}$, we
reconstruct the user's purchase sequence as it stood at
request time, $S_u^{<t}$, and append the partially masked
target step. The target step carries the request's H3 cell
$l$, calendar features derived from $t$, and the
domain-specific product-line mask
$\textsc{mask}^{\mathrm{dom}}$ (with
$\mathrm{dom} \in \{\text{rest}, \text{retail}\}$ chosen
according to which per-domain ranker is being trained). A
single forward pass of the trained classifier $h$ on this
input yields the unified logit vector, from which we gather
the per-candidate logits
$\{h_i(u, l, t)\}_{i \in \mathcal{I}_{u,l}^{t}}$ and append
$h_i(u, l, t)$ to $\mathbf{x}_{u,i,l}^{t}$ as one additional
ranker feature. Note that $h$ is invoked once per session,
not once per (session, candidate) pair: the candidate logits
are read out from the same forward pass via a single
$\mathcal{O}(|\mathcal{I}_{u,l}^{t}|)$ gather, so the offline
join scales linearly in the number of sessions rather than in
the number of (session, candidate) pairs.

\paragraph{Loss and labels.}
Both rankers use the session identifier as the ranking group. In V1 we
used a pairwise BPR-style logistic loss
(\textsc{PairLogit}) with the number of negatives
\texttt{max\_pairs} tuned per country, and negatives drawn
from impressed-but-not-purchased stores. With V2's
trial-oriented upgrade we switched to the list-aware
pairwise objective \textsc{YetiRankPairwise} with
$\mathrm{mode}\,{=}\,\mathrm{NDCG}$~\cite{gulin2011yetirank},
which let us additionally incorporate clicks as graded
relevance. Concretely, YetiRank builds a position-dependent weight
for each pair by repeatedly perturbing the current model scores
with random noise and re-ranking; pairs of candidates that land
adjacent near the top of these perturbed rankings accumulate a
larger weight than pairs deep in the list, concentrating learning
signal on the comparisons most likely to affect the top of the
served ranking~\cite{gulin2011yetirank}. Setting
$\mathrm{mode}\,{=}\,\mathrm{NDCG}$ selects NDCG as the target
metric this weighting is calibrated toward. We assign a per-row
label as a function of the purchase indicator $r_{u,i}^{t}$,
the click indicator $c_{u,i,l}^{t}$, and the user--store
relationship $\mathrm{rel}(u, i, t) \in \{\textsc{new}, \textsc{rec}\}$,
defined as \textsc{rec} if $i \in \mathcal{I}_u^{<t}$ and
\textsc{new} otherwise---the binary projection of
Equation~\ref{eq:purchase-type} extended to click rows, with
\textsc{cs} folded into \textsc{new}:
\begin{equation}
  y_{u,i,l}^{t} \;=\;
  \begin{cases}
    \lambda^{\,\textsc{p,new}} & \text{if } r_{u,i}^{t} = 1,\, \mathrm{rel} = \textsc{new}, \\
    \lambda^{\,\textsc{p,rec}} & \text{if } r_{u,i}^{t} = 1,\, \mathrm{rel} = \textsc{rec}, \\
    \lambda^{\,\textsc{c,new}} & \text{if } r_{u,i}^{t} = 0,\, c_{u,i,l}^{t} = 1,\, \mathrm{rel} = \textsc{new}, \\
    \lambda^{\,\textsc{c,rec}} & \text{if } r_{u,i}^{t} = 0,\, c_{u,i,l}^{t} = 1,\, \mathrm{rel} = \textsc{rec}, \\
    0                          & \text{otherwise.}
  \end{cases}
  \label{eq:ranker-label}
\end{equation}
The case predicates partition $\{0,1\}^{2}$ along the
purchase axis first, so a row that records both a click and a
purchase is treated as a purchase rather than a click. The
constants are calibrated so that purchase signal dominates
click signal and click signal in turn dominates the absence
of interaction:
$\lambda^{\,\textsc{p,new}} = \lambda^{\,\textsc{p,rec}} > \lambda^{\,\textsc{c,new}} > \lambda^{\,\textsc{c,rec}} > 0$.
The two purchase-row labels are equal because the
trial-vs-reorder distinction is already captured at the
session level via group weights, applied to the listwise
NDCG-style loss; the row label only needs to express the
relative strength of the implicit-feedback signal. We
additionally use per-session weights ordered
$w^{\textsc{new}} > w^{\textsc{rec}}$, biasing the ranker
toward trial sessions while preserving reorder performance.
Unlike the classifier, the ranker collapses cs and non-cs
new sessions into a single $w^{\textsc{new}}$; a separate
cs weight could yield finer control and is left to future work.

We frame this as a constrained multi-objective problem: maximize trial ranking quality subject to a bounded reorder regression. The per-session weight ratio ($w^{\textsc{new}}/w^{\textsc{rec}}$) is this constraint's empirically tuned Lagrange multiplier; Appendix~\ref{app:trial-reorder}'s grid sweep traces the resulting Pareto frontier as the multiplier varies. Tuning a single scalarization weight empirically, rather than deriving it analytically, mirrors established practice in multi-objective learning-to-rank and explore/exploit systems~\cite{chen2021values, su2024long}.

\paragraph{Per-domain rankers.}
We train one ranker for the restaurant domain (introduced in
V1) and, since V3, an additional ranker for the retail domain.
Both share the same loss and label scheme and most
optimization settings. Both are trained for $T_r = 500$
boosting iterations with tree depth $8$, learning rate $0.05$,
and L2 leaf regularization $3.0$, with early stopping on
$\mathcal{D}_{r}^{\text{val}}$ after $50$ rounds without
improvement.

\subsection{Testing and Benchmarking}
\label{sec:testing-benchmarking}

After training has concluded, we evaluate the per-domain ranker
on the held-out last day of sessions
$\mathcal{D}_{r}^{\text{test}}$. We compute MRR per purchase
type from Equation~\ref{eq:purchase-type}--separately for
\textsc{cs}, \textsc{new}, and \textsc{rec} targets--and across
four scoring functions: a benchmark $b$, the production ranking
$\rho^{\mathrm{prod}}$, the classifier $h$, and the trained
ranker $g$, yielding twelve MRR values per evaluation run and model domain. The
benchmark $b$ combines a binary reorder indicator with
a store-relative popularity score; aside from the
reorder indicator (which is itself user-specific) it uses
no further personalized signal, providing a strong yet
\emph{weakly personalized} reference. On the \textsc{new} slice
we consistently observe
\begin{equation}
  \mathrm{MRR}_{b} \;<\; \mathrm{MRR}_{\rho^{\mathrm{prod}}} \;<\; \mathrm{MRR}_{g}
  \quad\text{and}\quad \mathrm{MRR}_{h} \;<\; \mathrm{MRR}_{g},
  \label{eq:mrr-ordering}
\end{equation}
in every country and for both versions, while the position of
$h$ relative to $\rho^{\mathrm{prod}}$ is country-dependent
(Table~\ref{tab:offline-mrr}). On \textsc{rec} the ordering
depends on the operating point: at trial-weighted settings $g$
can fall below both $h$ and $\rho^{\mathrm{prod}}$
(Section~\ref{sec:offline}, Appendix~\ref{app:trial-reorder}).
We therefore apply the safeguard on the trial side: a new ranker
is blocked from roll-out if its MRR on the \textsc{new} segment
fails to exceed $\mathrm{MRR}_{b}$ at $95\%$ confidence (the
lower bound of a two-sigma interval around $\mathrm{MRR}_{g}$
must exceed $\mathrm{MRR}_{b}$). Performance on \textsc{rec} is
monitored as a guardrail and surfaces a warning, but does not
block deployment.

\section{Training and Serving Infrastructure}
\label{sec:infrastructure}
We launch parallel daily training jobs for each country on our internal Kubernetes cluster, orchestrated via Flyte\footnote{https://flyte.org/}. User interaction sequences for inference are generated during training and written to an online feature store to keep training and serving features synchronized. Since purchases are relatively sparse signals, daily updates were sufficient in our setting.

Additional feature groups--including content, contextual, and user-store interaction features--are computed through separate batch pipelines and published to the online feature store at cadences ranging from a few hours to daily, depending on upstream data availability.

At inference time, the ranking service retrieves all required features from the online feature store, computes the transformer-based classifier logits on CPUs, and applies the downstream GBDT ranker to produce the final store ranking. The end-to-end serving pipeline has a p99 latency of approximately 60 ms, including feature retrieval, model inference, and serialization.

\section{Experiment Results}
\label{sec:experiments}

We report offline and online results for V1 and V2, the two
production rollouts of UVR; their architectural and
loss-function differences are summarized in
Table~\ref{tab:uvr-versions}. V1 and V2 are restaurant-only; V3, our cross-domain
unification, is evaluated separately below
(Section~\ref{sec:online}). The offline benchmarks in this
section therefore focus on the restaurant domain, which
dominates Wolt's order volume. Substantial offline uplifts
gave us the conviction to carry both versions forward to
A/B~tests: V1 was rolled out in 2025 while V2 was deployed to production and replaced V1 in early 2026. V3 then replaced V2 a few months later. Throughout the section, we focus on a subset of our largest operating countries by purchase volume, which together
account for more than $50\%$ of Wolt's worldwide order volume and provide a representative cross-section while keeping
experimental complexity manageable.

\subsection{Baselines}
\label{sec:baselines}

We compare UVR against the production stack it was designed
to subsume, which consists of four ranking models split
across two domains. In the restaurant domain we operate
three models: Neural Collaborative Filtering~\cite{he2017ncf} with BPR-loss as a first-pass ranker (FPR), a DNN-based context- and content-aware second-pass ranker (SPR), and a GBDT-based ranker for cold-start users (CSR). In the retail domain we operate a single GBDT-based ranker (RR). The offline experiments below report uplifts of UVR over this stack, and the online A/B tests
(Section~\ref{sec:online}) use FPR+SPR/CSR as the V1 control
arm.

\subsection{Offline Testing}
\label{sec:offline}

We report the relative MRR improvement of the classifier $h$
and ranker $g$ over production, broken down by purchase type;
absolute volumes are omitted due to confidentiality. Headline
numbers are anchored at a Sunday peak-volume test day, with
results on other days directionally consistent.
Table~\ref{tab:offline-mrr} reports per-country uplifts for V1
and V2 on \textsc{new} and \textsc{rec}urring purchases; we
omit \textsc{cs} as its session volume in this slice is
negligible. A single-country study with
multiple training seeds shows run-to-run standard deviation of
$\sim$10-30\% relative to the mean uplifts; individual cells are therefore directional.

\begin{table}[t]
  \caption{Relative MRR uplift of V1/V2 over production, by country and purchase type (\textsc{new}/\textsc{rec}); classifier $h$ and ranker $g$ reported separately.}
  \label{tab:offline-mrr}
  \small
  \setlength{\tabcolsep}{4pt}
  \begin{tabular}{@{}llcccc@{}}
    \toprule
    & & \multicolumn{2}{c}{UVR V1} & \multicolumn{2}{c}{UVR V2} \\
    \cmidrule(lr){3-4}\cmidrule(lr){5-6}
    Country & Model & \textsc{new} & \textsc{rec} & \textsc{new} & \textsc{rec} \\
    \midrule
    \multirow{2}{*}{ISR} & Classifier & $-0.5\%$  & $-3.6\%$  & $+2.2\%$  & $\mathbf{-4.2\%}$  \\
                         & Ranker     & $\mathbf{+19.7\%}$ & $\mathbf{+5.0\%}$  & $\mathbf{+30.3\%}$ & $-12.5\%$ \\
    \midrule
    \multirow{2}{*}{GRC} & Classifier & $+4.7\%$  & $-8.7\%$  & $+5.0\%$  & $-8.7\%$  \\
                         & Ranker     & $\mathbf{+17.6\%}$ & $\mathbf{-5.0\%}$  & $\mathbf{+22.3\%}$ & $\mathbf{-7.8\%}$  \\
    \midrule
    \multirow{2}{*}{DEU} & Classifier & $+4.8\%$  & $-17.0\%$ & $+6.4\%$  & $\mathbf{-16.6\%}$ \\
                         & Ranker     & $\mathbf{+19.8\%}$ & $\mathbf{-11.4\%}$ & $\mathbf{+29.0\%}$ & $-27.3\%$ \\
    \midrule
    \multirow{2}{*}{FIN} & Classifier & $-4.8\%$  & $+6.2\%$  & $-3.5\%$  & $\mathbf{+7.7\%}$  \\
                         & Ranker     & $\mathbf{+7.5\%}$  & $\mathbf{+11.2\%}$ & $\mathbf{+12.5\%}$ & $+6.7\%$  \\
    \midrule
    \multirow{2}{*}{HUN} & Classifier & $-0.6\%$  & $+0.3\%$  & $-1.0\%$  & $\mathbf{+1.3\%}$  \\
                         & Ranker     & $\mathbf{+7.7\%}$  & $\mathbf{+9.4\%}$  & $\mathbf{+14.3\%}$ & $-1.4\%$  \\
    \midrule
    \multirow{2}{*}{DNK} & Classifier & $-3.1\%$  & $+4.0\%$  & $-1.7\%$  & $\mathbf{+5.7\%}$  \\
                         & Ranker     & $\mathbf{+9.0\%}$  & $\mathbf{+7.6\%}$  & $\mathbf{+14.5\%}$ & $-5.8\%$  \\
    \bottomrule
  \end{tabular}
\end{table}

\paragraph{V1 (PairLogit) over the production ranker.}
The ranker $g$ improves over the classifier $h$ in all
$12$ of $12$ (country, purchase-type) cells, confirming that
content and contextual features carry information
complementary to the sequence signal alone. The improvement
holds uniformly across both \textsc{new} and \textsc{rec}
purchases: every country gains on \textsc{new} (range
$+7.5\%$ to $+19.8\%$), and the ranker also recovers
ground on \textsc{rec} relative to the classifier in every
country.

\paragraph{V2 (YetiRankPairwise:NDCG) over V1.}
The picture splits along the trial/reorder axis. On
\textsc{new}, V2 amplifies the V1 gains uniformly across
all six countries (e.g.\ ISR $+19.7\%\!\rightarrow\!+30.3\%$,
DEU $+19.8\%\!\rightarrow\!+29.0\%$), in line with the
design intent of the listwise NDCG-aware loss. On
\textsc{rec}urring purchases, however, the classifier
alone outperforms the V2 ranker in five of six countries
(ISR, DEU, FIN, HUN, DNK; only GRC remains in the ranker's
favor), suggesting that the trial-oriented loss
penalizes reorder relevance more aggressively than V1's
PairLogit. Section~\ref{sec:online} tests whether this trade-off surfaces in production; Appendix~\ref{app:trial-reorder}
dissects the ranker-level levers driving it via a controlled
ablation with the classifier held fixed.

\paragraph{Operating point selection.}
From the sweep of Appendix~\ref{app:trial-reorder} we discard every
setting with a negative \textsc{new} MRR uplift, then take the
setting with the largest \textsc{new} uplift whose \textsc{rec}
regression stays within $-15\%$. This makes the constraint of
Section~\ref{sec:training-ranker} explicit. DEU is the one country where no setting satisfied the bound.

\subsection{Online A/B Testing}
\label{sec:online}

The offline uplifts in Section~\ref{sec:offline}, combined
with our internal experience with the predecessor FPR and SPR
rankers (where offline MRR gains had consistently translated
into online wins), motivated three consecutive A/B~tests.
All three tests used a 50:50 user-level random split over all
registered users, cold-start and returning alike. V1 and V2 ran
for two weeks across the ten largest operating countries; V3 ran
for three weeks across all operating countries, reflecting the
larger sample needed for retail-domain significance. The target
metrics were the daily global conversion rate (Global CVR) and
the daily Merchant Trial Rate, which captures first-time ordering
from a store. Table~\ref{tab:online-ab} summarizes the
per-version outcomes.

\begin{table}[t]
  \caption{Online A/B outcomes per UVR version (V1 vs.\ FPR+SPR/CSR;
  V2 vs.\ V1; V3 vs.\ V2+RR). Merchant Trial Rate gains are significant
  at $\alpha = 0.05$; Global CVR is reported with its $p$-value.}
  \label{tab:online-ab}
  \small
  \begin{tabular}{@{}lcc@{}}
    \toprule
    Version & Merchant Trial Rate & Global CVR \\
    \midrule
    V1 & $+5.5\%$           & $+0.16\%$ ($p = 0.017$) \\
    V2 & $+0.45\%$          & $-0.05\%$ ($p = 0.398$) \\
    V3 & $+1.31\%^{\dagger}$   & $+0.04\%$ ($p = 0.457$) \\
    \bottomrule
    \addlinespace[2pt]
    \multicolumn{3}{@{}l@{}}{\footnotesize $^{\dagger}$Retail domain.} \\
  \end{tabular}
\end{table}

Three observations stand out. First, the V1 Merchant Trial Rate
gain of $+5.5\%$ confirms that UVR effectively pushes
trial in production; combined with the
simultaneous $+0.16\%$ Global CVR uplift, this shows that the
trial push does not come at the expense of conversions. Second, V2's $+0.45\%$
Merchant Trial Rate gain on top of an already strong V1
baseline confirms that the listwise loss, along with label smoothing and an additional trial feature, continues to deliver
incremental trial value.
Third, V3's $+1.31\%$ Merchant Trial Rate gain is measured on
the retail domain--V1 and V2's figures above are
restaurant-domain--confirming that unifying the restaurant
and retail rankers lifts retail discovery with no significant
change in restaurant-domain trial rate or blended Global CVR. Ads revenue regressed significantly under both V1 and V3
($-0.43\%$ for V3), which we read as confirmation of
effectiveness: personalized ranking outgrew the Ads stack, which
still relied on a variant of the previous production restaurant
ranker. RR was retired upon V3's rollout, completing the
consolidation of four legacy rankers into one.

\subsection{Discussion and Limitations}
\label{sec:discussion}

The online results confirm that the trial side of the trade-off can be
steered effectively, and that steering it is compatible with a conversion
gain (V1: $+5.5\%$ Merchant Trial Rate with $+0.16\%$ Global CVR). The partial offline reorder regressions of Section~\ref{sec:offline}
left the blended metric statistically unchanged ($|\Delta| < 0.1\%$, $p > 0.39$). Beyond ranking quality, consolidating FPR, SPR, CSR, and RR
into a single cross-domain model (V3) substantially reduces operational
complexity, retiring the last of four legacy rankers. The control arm
reflects the original rank, which includes ad placements and
deduplication beyond the ranker, so a direct model-to-model comparison
would likely narrow the observed gap.

The V3 result validates the business case for cross-domain
unification but does not by itself isolate the contribution of
cross-domain \emph{transfer}, i.e., whether retail
benefits from shared restaurant purchase signal, or simply from
finally having a transformer+GBDT stack rather than a single GBDT
(RR). Two ablations would sharpen the
attribution: a domain-siloed transformer+GBDT retail model,
separating transfer from the mere presence of a sequence stage,
and an otherwise-identical ranker trained without the classifier
logit, isolating the transformer's marginal contribution. The
closest evidence we report is the classifier-versus-ranker
comparison of Table~\ref{tab:offline-mrr} alongside the two
purely tree-based production baselines (CSR, RR); feature sets, losses, and training data differ, so neither is controlled.
Under a fixed engineering timeline we prioritized the online
tests; each ablation is a single training run and we will report
both.

Three limitations are worth noting. First, UVR runs on purchase
histories with up to one day of delay and does not exploit real-time
interaction signals; the architecture could readily accommodate these
as first-class sequence events. Second, the transductive prediction
head restricts the classifier to stores seen during training, leaving
new stores reliant on the GBDT's content features alone. Third,
restricting the sequence to purchases leaves other useful signals (clicks, browsing,
dwell time) untapped, which could benefit users with sparse histories.

\section{Conclusion}
\label{sec:conclusion}

We presented UVR, a production transformer--GBDT system that
consolidates four specialized rankers---three for restaurants, one
for retail---into a single unified model. Three consecutive A/B
tests validated the approach: V1 delivered $+5.5\%$ Merchant Trial
Rate with a simultaneous $+0.16\%$ Global CVR uplift; V2 added a
further $+0.45\%$ Merchant Trial Rate; and V3, our cross-domain
unification, delivered a further $+1.31\%$ Retail Merchant Trial
Rate while retiring the last domain-specific legacy ranker. The
hybrid architecture makes the trial/reorder balance an explicit
operating point: the per-session weight ratio
$w^{\textsc{new}}/w^{\textsc{rec}}$ and the sampled-negative count
\texttt{max\_pairs} scalarize a constrained multi-objective problem,
and 150 ranker retrains trace the frontier they span
(Appendix~\ref{app:trial-reorder}). At the ratio we deploy, this buys
$+12\%$ to $+30\%$ offline trial MRR over production and costs
reorder MRR in five of six countries, a trade that did not surface in
Global CVR. Future work focuses on target-aware conditioning of the transformer
on the candidate stores, inductive content store representations for
cold-start mitigation, and real-time interaction signals for short-term intent.

\begin{acks}
The authors would like to thank Adam Smaili, Daniel Lazar, Paavo Camps, Tanja Menković, Ibrahim Omer Celik, Dzung Nguyen, Sergio Gonzalez Sanz as well as the Machine Learning Platform, Analytics and Product Management teams at Wolt who contributed or supported us throughout this project.
\end{acks}

\FloatBarrier

\clearpage
\bibliographystyle{ACM-Reference-Format}
\bibliography{references}

\clearpage
\appendix
\raggedbottom

\section{Trial vs. Reorder Trade-off}
\label{app:trial-reorder}

\begin{figure}[H]
  \centering
  \includegraphics[width=\columnwidth]{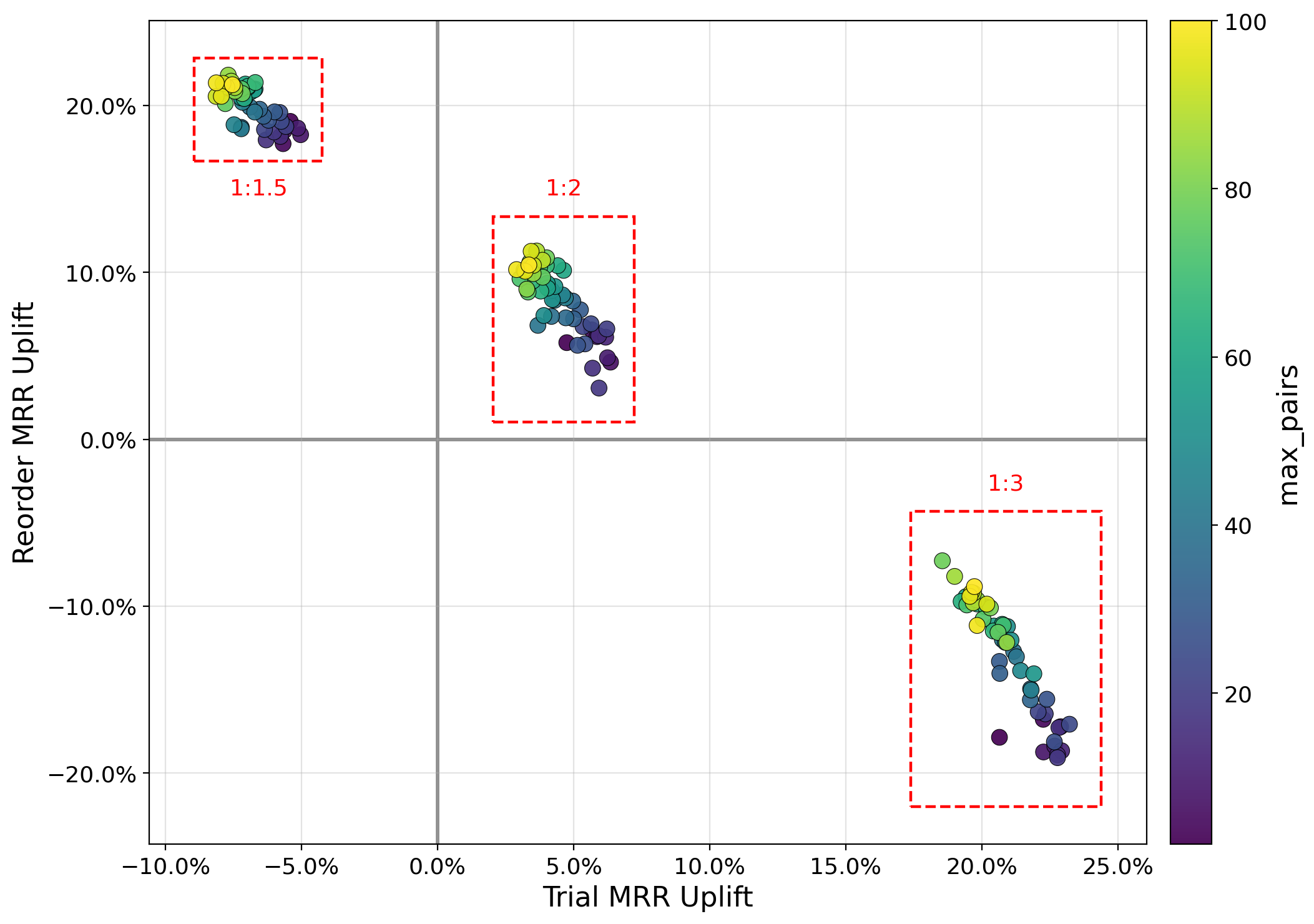}
  \caption{Ranker ablation: relative MRR uplift over
  production on \textsc{new} (x) vs.\ \textsc{rec} (y) with the
  classifier held fixed.}
  \label{fig:app-tradeoff}
\end{figure}

To isolate the ranker's contribution to the trial vs.\ reorder
trade-off, we hold the classifier fixed and re-train only the
ranker across 3 per-session weighting regimes
$w^{\textsc{new}}/w^{\textsc{rec}} \in \{1.5, 2.0, 3.0\}$
($w^{\textsc{rec}}{=}1$; the three clusters in
Figure~\ref{fig:app-tradeoff}) and 50 \texttt{max\_pairs}
negative-sampling settings per regime (dot color). All 150 runs
share the same country, day, and random seed.

Axes report relative MRR uplift over the production baseline on
\textsc{new} (x) and \textsc{rec} (y) purchases. The clusters
trace a Pareto-style frontier:
$w^{\textsc{new}}/w^{\textsc{rec}}{=}1.5$ yields strong reorder
gains with a small trial regression, $2.0$ balances both at
modest positive uplifts, and $3.0$ delivers a strong trial
uplift at the cost of reorder. Within each cluster, increasing
\texttt{max\_pairs} (low~$\to$~high) consistently shifts the
model results toward stronger reorder and weaker
trial. Together these two ranker-only levers enabled us to tune the trade-off between trial and reorder.

\section{MRR vs. Maximum Sequence Length}
\label{app:user-sequence-analysis}

\begin{figure}[H]
  \centering
  \includegraphics[width=0.9\columnwidth]{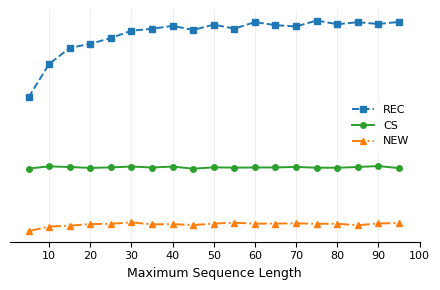}
  \caption{Classifier MRR on the \textsc{rec}, \textsc{cs},
  and \textsc{new} validation slices vs.\ the maximum input
  sequence length.}
  \label{fig:app-seqlen}
\end{figure}

Figure~\ref{fig:app-seqlen} shows that reorder MRR rises sharply
up to $\sim$30 purchases and then plateaus, while trial and
cold-start MRR are largely insensitive to
sequence length. This motivates the moderate sequence cap $M$
used in training (Section~\ref{sec:training-classifier}).

\end{document}